\documentclass[10pt]{IEEEtran}
\usepackage[T1]{fontenc}
\usepackage{amsmath,graphicx,amssymb,dsfont,cite}
\usepackage{float}
\usepackage{caption}
\usepackage{epsfig}
\usepackage{algorithm,algorithmicx,algpseudocode}
\usepackage{color}
\usepackage{xcolor}
\usepackage{soul}

\newtheorem{proposition}{Proposition}

\title{Semi-independent resampling for particle filtering}
         
\author{Roland Lamberti, Yohan Petetin, Fran\c cois Desbouvries, \IEEEmembership{Senior Member,~IEEE,} and Fran\c cois Septier 
\thanks{
R. Lamberti, Y. Petetin and F. Desbouvries are with Samovar, Telecom Sudparis, CNRS, Universit\'e Paris-Saclay, 9 rue Charles Fourier, 91011 Evry, France.
F. Septier is with IMT Lille Douai, Univ. Lille, CNRS, UMR 9189 - CRIStAL, F-59000 Lille, France.}
}

\begin{document}
\maketitle

\begin{abstract}

Among Sequential Monte Carlo (SMC) methods,
Sampling Importance Resampling (SIR) algorithms 
are based on Importance Sampling (IS) and on some (resampling-based)
rejuvenation algorithm which aims at fighting against weight degeneracy.
However 
this mechanism tends to be insufficient when applied to informative or high-dimensional models. 
In this paper we revisit the rejuvenation mechanism 
and propose a class of parameterized SIR-based solutions 
which enable to adjust the tradeoff between computational cost and statistical performances.
\end{abstract}

\section{Introduction and background}
\label{intro}
Bayesian filtering consists in estimating some variable $x_t$ from noisy measurements $y_{0:t}=\{y_0,\cdots,y_t\}$.
We assume that $\{(x_t,y_t)\}_{t \geq 0}$ is a Hidden Markov Chain, i.e. that the joint density of $(x_{0:t},y_{0:t})$ reads
$
p(x_{0:t},y_{0:t}) = p(x_{0})\prod_{s = 1}^{t} f_{s}(x_{s}|x_{s - 1}) \prod_{s = 0}^{t} g_{s}(y_{s}|x_{s})\text{.}
$
The problem can be traced back to Kalman \cite{kalman}
in the context of linear and Gaussian state space models.
Approximate solutions for non linear and/or non Gaussian state space models include 
the extended Kalman filter
\cite{jazwinski, andersonmoore, Ristic-Kalman },
the unscented Kalman filter 
\cite{julier-procieee, tanizaki, arulampalam, systematization-UKF},
or SMC methods
(also called particle filters (PF))
\cite{gordon-salmond-smith,livredoucet,arulampalam},
which propagate in time a discrete approximation 
$\widehat{p}(x_t|y_{0:t})=\sum_{i=1}^N w_t^i \delta_{x_t^i}$
of the posterior pdf $p(x_t|y_{0:t})$.

\subsection{The classical SIR algorithm}
\label{classical-sir}

Let $\Theta_t = \int \varphi(x_t) p(x_t|y_{0:t}) {\rm d}x_t$
be a moment of interest of $p(x_t|y_{0:t}) $.
One iteration of an SMC algorithm can be decomposed in three steps.

Starting at time $t-1$ from 
$\{w_{t-1}^i,x_{t-1}^i\}_{i=1}^N$,
the first two steps consist in sampling (\textsl{S.}) $N$ particles $\tilde{x}_t^i$
from importance densities 
$q_i$
and weighting (\textsl{W.}) them so as to take into account 
the discrepancy between the target and importance densities;
then $\Theta_t$ is estimated as
$\widehat{\Theta}_t^{{\rm SIS},N} = \sum_{i=1}^N \tilde{w}_t^i \varphi(\tilde{x}_t^i)$
(superscript SIS will be justified below).
Finally a third (optional) step consists in re-sampling (\textsl{R.}) the weighted particles,
i.e. in re-drawing each particle with a probability equal to its weight
and assigning to the resampled particles the same weight $\frac{1}{N}$.
This yields the class of
SIR algorithms
\cite{smith-gelfand} 
\cite{gordon-salmond-smith}
\cite{livredoucet}
\cite{arulampalam}
described by Algorithm \ref{algo-PFSIR}.

\begin{algorithm}
\caption{The classical SIR algorithm}
\label{algo-PFSIR}
\begin{algorithmic}
\State \textbf{Data:} $q(x_t|x_{t-1})$, $y_t$, $\{w_{t-1}^i,x_{t-1}^i\}_{i=1}^{N}$
\For {$1 \leq i \leq N$} 
\State \textbf{S.} $\tilde{x}_t^{i} \sim q(x_t|x_{t-1}^i)$;  
\State \textbf{W.} $\tilde{w}_t^{i} \propto  w_{t-1}^i\frac{f_{t}(\tilde{x}_t^i|x_{t-1}^i)g_t(y_t|\tilde{x}_t^i)}{q(\tilde{x}_t^i|x_{t-1}^i)}$ \text{, } $\sum_{i=1}^N \tilde{w}_t^i =1$;
\EndFor
\State $\widehat{\Theta}_t^{{\rm SIS},N} = \sum_{i=1}^N \tilde{w}_t^i \varphi(\tilde{x}_t^i)$;
\If {\textbf{R.}}
\For{$1 \leq i \leq N$} 
\State  $l^i \sim {\rm Pr}(L=l) = \tilde{w}_t^l,$ $1 \leq l \leq N$; 
\EndFor
\State  Set $\{w_t^i, x_t^i\}_{i=1}^N = \{\frac{1}{N} , \tilde{x}_t^{l_i}\}_{i=1}^N$.
\Else
\State  Set $\{w_t^i, x_t^i\}_{i=1}^N = \{\tilde{w}_t^i , \tilde{x}_t^{i}\}_{i=1}^N$.
\EndIf
\end{algorithmic}
\end{algorithm}

Let us comment 
this algorithm.
If resampling is totally absent, 
each time iteration 
reduces to the first two steps, i.e. is based on IS only.
However such a sequential IS (SIS) algorithm is well known to fail in practice
since after a few iterations most weights get close to zero.
The third step 
(which can be performed whatever $t$
or depending on some criterion such as the number of effective particles
\cite{kong1994} 
\cite{liu-chen1995}
\cite{liu1996metropolized}
\cite{cornebise})
discards particles with low weights 
(such particles are likely to be never resampled)
and is considered as a traditional rescue against weight degeneracy.
On the other hand, this (\textsl{R.}) step introduces local extra variance 
\cite[section 4.2.1]{liu-chen1995},
\cite[p. 213]{cappe2005},
which in turn affects the variance of $\widehat{\Theta}_t^{{\rm SIS},N}$ at subsequent iterations.
It has thus been proposed to control this extra variance term via
alternative resampling schemes
(see e.g. \cite {kitagawa1996} \cite{douc-cappe-resampling} \cite{Li:2015fl} $\cdots$).
Yet despite many proposed refinements
this generic SIR mechanism remains inefficient in informative models featuring very sharp likelihood functions
(i.e., when $g_t(y_t|x_t)$ is very small for most values of $x_{t}$), 
and in particular in high-dimensional state-space models \cite{Snyder-PF,Snyder2011}.

\subsection{The independent SIR algorithm}

Recently it has thus been proposed to revisit the SIR algorithm
\cite{icassp2016}
\cite{ssp2016}
\cite{ieeetsp-independent}
and more precisely to come back to the rejuvenation mechanism (\textsl{R.}).
The counterpart of this (\textsl{R.}) step is that it duplicates particles with high weights, 
which results in support degeneracy.
Moreover given $\{ w_{t-1}^i, x_{t-1}^i \}_{i=1}^N$
the samples $\{ x_t^j \}$ produced by Algorithm \ref{algo-PFSIR} 
are 
marginally distributed from some compound pdf $\tilde{q}_t^N$
which takes into account the effects of 
the three elementary (\textsl{S.}), (\textsl{W.}) and (\textsl{R.}) steps,
but are obviousy dependent \cite{icassp2016}
(a single particle can be re-sampled more than once);
by contrast, given $\{ w_{t-1}^i, x_{t-1}^i \}_{i=1}^N$
the independent SIR Algorithm \cite{icassp2016} \cite{ieeetsp-independent}
produces $N$ i.i.d. draws from $\tilde{q}_t^N$.
Note that Algorithm \ref{algo-PFSIR-ind} below only decribes the rejuvenation step of the independent SIR algorithm, 
and replaces the "{\bf if R. then}" part of Algorithm \ref{algo-PFSIR}.

\begin{algorithm}
\caption{Indep. SIR algorithm ({\sl resampling step only})}
\label{algo-PFSIR-ind}
\begin{algorithmic}
\State \textbf{Data:} $q(x_t|x_{t-1})$, $y_t$, $\{w_{t-1}^i,x_{t-1}^i\}_{i=1}^{N}$;
\For{$1 \leq j \leq N$} 
\State
$\tilde{x}_t^{1,j} \leftarrow \tilde{x}_t^{j}$,
$\tilde{w}_t^{1,j} \leftarrow \tilde{w}_t^{j}$.
\EndFor
\For{$1 \leq i \leq N$} 
\State \textbf{R.} 
$l^i \sim {\rm Pr}(L = l) =
\tilde{w}_t^{i,l}$, $1\leq l \leq N$;
\State {\sl Rejuvenation of the support for iteration} $i+1$
  \If{$(i < N)$}
\For{$1 \leq j \leq N$} 
\State $\tilde{x_t}^{i+1,j} \sim q(x_t|x_{t-1}^{j})$;  
\State $\overline{w}_t^{i+1,j} 
=
w_{t-1}^{j} 
\frac{ f_{t}(\tilde{x}_t^{i+1,j}|x_{t-1}^{j})  g_t(y_t|\tilde{x}_t^{i+1,j})}{q(\tilde{x}_t^{i+1,j}|x_{t-1}^{j})},$
\EndFor
\State $\tilde{w}^{i+1,:}_{t} \propto \overline{w}^{i+1,:}_{t}$, $\sum^{N}_{j = 1}{\tilde{w}^{i+1,j}_{t}} = 1$;
\EndIf
\EndFor
\State  Set $\{ w_t^i, x_t^i \}_{i=1}^N = \{ \frac{1}{N} , \tilde{x}_t^{i,l^i} \}_{i=1}^N$.
\end{algorithmic}
\end{algorithm}

\subsection{Scope of the paper}

Algorithm \ref{algo-PFSIR-ind} has displayed good results in severe situations \cite{icassp2016}
and can be combined with a post-resampling, second-stage reweigthing scheme due to its 
auxiliary particle filtering interpretation \cite{ssp2016} \cite{ieeetsp-independent}.
However its rejuvenation mechanism involves the sampling 
of $N^2$ particles $\{ \tilde{x_t}^{i,j} \}_{i,j=1}^N$
(plus $N$ resampling steps),
by contrast with the classical SIR algorithm which only samples 
$N$ intermediate particles $\{ \tilde{x}_t^{j} \}_{j=1}^N$
(and is also followed by $N$ resampling steps).
One can wonder whether this extra cost is indeed necessary,
so the aim of this paper 
is to design an algorithm which is both 
efficient (in terms of computational cost) and
effective (in terms of statistical results).
The rest of this paper is organized as follows.
Our algorithm is described in section \ref{semiind}.
Simulations are displayed in section \ref{sec:simu}, 
and the paper ends with a conclusion.

\section{Semi-independent resampling}
\label{semiind}

\subsection{An intermediate resampling scheme}
\label{intermediate}

\begin{figure}[htbp!]
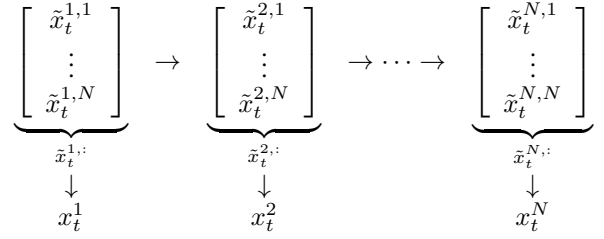

\centering
$$
\begin{array}{ccccc}
\underbrace{
\left[
\begin{array}{c}
\tilde{x}_{t}^{1,1} \\
\vdots \\
\tilde{x}_{t}^{1,N}
\end{array}
\right]
}_{\tilde{x}_{t}^{1,:}}
&
\rightarrow
&
\underbrace{
\left[
\begin{array}{c}
\tilde{x}_{t}^{2,1} \\
\vdots \\
\tilde{x}_{t}^{2,N}
\end{array}
\right]
}_{\tilde{x}_{t}^{2,:}}
&
\rightarrow \cdots \rightarrow
&
\underbrace{
\left[
\begin{array}{c}
\tilde{x}_{t}^{N,1} \\
\vdots \\
\tilde{x}_{t}^{N,N}
\end{array}
\right]
}_{\tilde{x}_{t}^{N,:}}
\\
\downarrow
&
&
\downarrow
&
&
\downarrow
\\
x_t^1
&
&
x_t^2
&
&
x_t^N
\end{array}
$$
\caption{\small The classical, independent and semi-independent resampling mechanisms. 
Each scheme draws $N$ supports $\tilde{x}_{t}^{i,:}$ 
and redraws one sample $x_t^i$ out of each support. 
The difference lies in the way $\tilde{x}_{t}^{i,:}$ is built from $\tilde{x}_{t}^{i-1,:}$:
$\tilde{x}_{t}^{i,j}$ is a copy of $\tilde{x}_{t}^{i-1,j}$ in the classical case;
is a new particle in the independent case;
and can be either copied of redrawn in the intermediate, semi-independent case.} 
\label{fig:slices}
\end{figure}

The classical and independent SIR resampling mechanisms 
can be reconciled in a common framework.
In both schemes, one progressively builds 
$N$ weighted sets 
$\tilde{x}_{t}^{1,:}$, $\cdots$, $\tilde{x}_{t}^{N,:}$
(the $N$ supports) 
and redraws one sample $x_t^i$ out of each of them
(see figure \ref{fig:slices}). 
The difference lies in the way $\tilde{x}_{t}^{i,:}$ is built from $\tilde{x}_{t}^{i-1,:}$:
in the classical SIR mechanism, 
$\tilde{x}_{t}^{i,:}$ is a copy of 
$\tilde{x}_{t}^{i-1,:}$
(so the resampling step amounts to redrawing $N$ samples from the common support 
$\tilde{x}_{t}^{1,:}$, see Algorithm \ref{algo-PFSIR}); 
in the independent SIR mechanism, 
a whole new support $\tilde{x}_{t}^{i,:}$ 
is drawn at each iteration $i$. 
In other words, 
from a computational point of view
both schemes resample $N$ particles from some intermediate set $\{ \tilde{x}_{t}^{i,j} \}_{i,j=1}^N$,
but building that set requires 
$N$ preliminary independent sampling steps in the classical case, 
while it requires $N^2$ independent sampling steps in the independent case.

In this paper we propose a resampling scheme which creates an intermediate set $\{ \tilde{x}_{t}^{i,j} \}_{i,j=1}^N$
with more diversity than in the classical case, 
but at a reduced sampling cost as compared to the independent case.
Starting from $\tilde{x}_{t}^{i-1,j}$, 
$\tilde{x}_{t}^{i,j}$ can now either be a copy (to save cost) 
or a new sample (to enhance diversity).
The algorithm is as follows. 
Fix the number $k$ (with $0\leq k \leq N$) of samples which will be redrawn at each iteration.
At step $i$,
uniformly draw a subset $m^{i,1:k} = (m^{i,1},\cdots,m^{i,k})$ of size $k$ out of $(1,\cdots,N)$
(
$m^{i,l}$ are the indices of the particles which will be redrawn).
Next 
$\tilde{x}_{t}^{i,j} \sim q(x_t|x_{t-1}^j)$ if $j \in m^{i,1:k}$, 
and
$\tilde{x}_{t}^{i,j} = \tilde{x}_{t}^{i-1,j}$ if $j \notin m^{i,1:k}$.
Finally observe that the classical (resp. independent) SIR algorithm corresponds to the particular case $k=0$ (resp. $k=N$).
The algorithm is summarized in Algorithm \ref{algo-PFSR} below.

\begin{algorithm}
\caption{Semi-ind. SIR algorithm ({\sl resampling step only})}
\label{algo-PFSR}
\begin{algorithmic}
\State \textbf{Data:} $q(x_t|x_{t-1})$, $y_t$, $\{w_{t-1}^i,x_{t-1}^i\}_{i=1}^{N}$
\For{$1 \leq j \leq N$} 
\State
$\tilde{x}_t^{1,j} \leftarrow \tilde{x}_t^{j}$,
$\tilde{w}_t^{1,j} \leftarrow \tilde{w}_t^{j}$.
\EndFor
\For{$1 \leq i \leq N$} 
\State \textbf{R.} 
$l^i \! \sim \! {\rm Pr}(L \!=\! l)= \tilde{w}_t^{i,l}$, $1 \leq l \leq N$;
\State \textit{Partial rejuvenation of the support for iteration} $i+1$
  \If{$(i < N)$}
	\State $\tilde{x}_t^{i+1,:} \leftarrow \tilde{x}_t^{i,:}$, $\overline{w}_t^{i+1,:} \leftarrow \overline{w}_t^{i,:}$;
  \For{$1 \leq j \leq k$}
\State \mbox{$m^j \sim {\rm Pr}(M = n| n \in 1 \mathalpha{:} N \textbackslash \{m^{1:j - 1}\})= \frac{1}{N - j + 1}$;}  
\State 
$\tilde{x}^{i+1,m^{j}}_{t} \sim q(x_{t}|x^{m^{j}}_{t - 1})$;
\State 
$\overline{w}^{i+1,m^{j}}_{t} = w^{m^{j}}_{t - 1}
\frac{f_{t}(\tilde{x}^{i+1,m^{j}}_{t}|x^{m^{j}}_{t - 1})g_{t}(y_{t}|\tilde{x}^{i+1,m^{j}}_{t})}{q(\tilde{x}^{i+1,m^{j}}_{t}|x^{m^{j}}_{t - 1})}$;
  \EndFor
\State $\tilde{w}^{i+1,:}_{t} \propto \overline{w}^{i+1,:}_{t}$, $\sum^{N}_{j = 1}{\tilde{w}^{i+1,j}_{t}} = 1$;
\EndIf
\EndFor
\State  Set $\{w_t^i, x_t^i\}_{i=1}^N = \{\frac{1}{N} , \tilde{x}_t^{i,l^i}\}_{i=1}^N$.
\end{algorithmic}
\end{algorithm}

\subsection{Performances vs. computational cost}
\label{compromise}

We now evaluate the performance of this procedure 
by comparing the variances of the estimates computed
after the resampling step because they affect 
the variances of the estimates at subsequent iterations  \cite{douc-cappe-resampling}.
So let 
$\widehat{\Theta}_t^{{\bf .},N}  = \frac{1}{N} \sum_{i=1}^N \varphi(x_t^{{\bf .}, i})$,
where the generic notation $x_t^{{\bf .}, i}$ represents the points produced either by Algorithm 
\ref{algo-PFSIR},
\ref{algo-PFSIR-ind} or
\ref{algo-PFSR}
(so we consider $\widehat{\Theta}_t^{{\rm SIR},N}$,
$\widehat{\Theta}_t^{{\rm I-SIR},N}$ and
$\widehat{\Theta}_t^{{\rm SR},N,k}$, where SR stands for
semi-resampling).
We have the following proposition (the proof is given in the Appendix).

\begin{proposition}
\label{prop:1}
Given the previous set of particles $\{x_{0:t-1}^i\}_{i=1}^N$,
for all $k$, $0 \leq k \leq N$, we have: 
\begin{subequations}
\begin{align}
 {\rm E}(\widehat{\Theta}_t^{{\rm SR},N,k})&= {\rm E}(\widehat{\Theta}_t^{{\rm I-SIR},N})= {\rm E}(\widehat{\Theta}_t^{{\rm SIR},N}) \text{,} \label{prop:1:a} \\
 {\rm var}(\widehat{\Theta}_t^{{\rm I-SIR},N}) &\leq {\rm var}(\widehat{\Theta}_t^{{\rm SR},N,k}) \leq {\rm var}(\widehat{\Theta}_t^{{\rm SIR},N})  \text{,} \label{prop:1:b} \\
 {\rm var}(\widehat{\Theta}_t^{{\rm SR},N,k})& \leq {\rm var}(\widehat{\Theta}_t^{{\rm SR},N,k - 1}) \text{.} \label{prop:1:c}
\end{align}
\end{subequations}
\end{proposition}
So as the number $k$ of intermediate redrawings increases from $0$ to $N$,
the conditional variance 
of the semi-independent resampling estimator $\widehat{{\Theta}}_t^{{\rm SR},N,k}$ 
decreases from the upper bound of inequality \eqref{prop:1:b}
(if $k=0$, $\widehat{{\Theta}}_t^{{\rm SR},N,0}$ reduces to $\widehat{{\Theta}}_t^{{\rm SIR},N}$)
to its lower bound  
(if $k=N$, ${\Theta}_t^{{\rm SR},N,N}$ reduces to $\widehat{{\Theta}}_t^{{\rm I-SIR},N}$).
However remember from section \ref{intermediate} 
that $N+(N-1)\times k$ samples are needed for building 
$\widehat{{\Theta}}_t^{{\rm SR},N,k}$;
so parameter $k$ of the 
SR scheme 
enables to fix a compromise between variance reduction and computational budget.

\subsection{A parallelized version}
\label{parallel}

Finally Algorithm \ref{algo-PFSR} can be transformed into
a parallelized version,
the non-sequential SR (NSSR) algorithm.
At iteration $i$, 
instead of duplicating the $N-k$ surviving particles 
from the previous support $\tilde{x}_{t}^{i-1,:}$ 
(see Fig. \ref{fig:slices}),
we propose to duplicate the $N-k$ surviving particles 
directly from the initial set $\tilde{x}_{t}^{1,:}$ of particles. 
The $N-1$ new supports can thus be produced in parallel,
contrary to Algorithm \ref{algo-PFSR} which by nature is sequential.
Of course, this procedure alters the diversity of the final set of particles,
as is illustrated by the following proposition.

\begin{proposition}
\label{prop:2}
Let $\widehat\Theta^{NSSR,k}_{t}$ be the estimate built 
from the non-sequential semi-independent resampling procedure. 
Then given the previous set of particles $\{x_{0:t-1}^i\}_{i=1}^N$,
for all $k$, $0 \leq k \leq N$, we have:

\begin{subequations}
\begin{align}
 {\rm E}(\widehat{\Theta}_t^{{\rm NSSR},N,k})&= {\rm E}(\widehat{\Theta}_t^{{\rm I-SIR},N})= {\rm E}(\widehat{\Theta}_t^{{\rm SIR},N}) \text{,} \label{prop:2:a} \\
 {\rm var}(\widehat{\Theta}_t^{{\rm I-SIR},N}) &\leq {\rm var}(\widehat{\Theta}_t^{{\rm NSSR},N,k}) \leq {\rm var}(\widehat{\Theta}_t^{{\rm SIR},N})  \text{,} \label{prop:2:b} \\
{\rm var}(\widehat{\Theta}_t^{{\rm NSSR},N,k})& \leq {\rm var}(\widehat{\Theta}_t^{{\rm NSSR},N,k - 1}) \text{,} \label{prop:2:c} \\
{\rm var}(\widehat{\Theta}_t^{{\rm SR},N,k})& \leq {\rm var}(\widehat{\Theta}_t^{{\rm NSSR},N,k}) \text{.} \label{prop:2:d}
\end{align}
\end{subequations}
\end{proposition}
So we see that 
${\rm var}(\widehat{\Theta}_t^{{\rm NSSR},N,k})$ still decreases with $k$,
but is always larger than ${\rm var}(\widehat{\Theta}_t^{{\rm SR},N,k})$.
As with Proposition \ref{prop:1},
the variance inequalities still rely on Jensen's inequality, 
and the proof is omitted.

\section{Simulations}
\label{sec:simu}

We consider a tracking problem based on range-bearing measurements.
The hidden state-vector contains the position and velocity of the target in cartesian coordinates,
$x_{t} = [c_{x,t}, \dot{c}_{x,t}, c_{y,t}, \dot{c}_{y,t}]^{T}$. 
We set 
$f_t(x_t|x_{t-1})= \mathcal{N}(x_{t};{\bf F}x_{t - 1}; {\bf Q})$,
$g_t(y_{t}|x_{t}) = \mathcal{N}(y_{t};$
$\begin{bmatrix}
\sqrt{c^{2}_{x,t} + c^{2}_{y,t}};
\arctan{\frac{c_{y,t}}{c_{x,t}}}
\end{bmatrix}^T;{\bf R})$,
with 
$\mathbf{R}= {\rm diag}
(\sigma^{2}_{\rho}, \sigma^{2}_{\theta})$,
${\bf F} =
{\bf I}_2 
\otimes
\begin{bmatrix}
1 & 1 \\
0 & 1  
\end{bmatrix}$,
$\mathbf{Q} = 10 \times
{\bf I}_2 
\otimes 
\begin{bmatrix}
\frac{1}{3} & \frac{1}{2} \\
\frac{1}{2} & 1  
\end{bmatrix}$
where $\otimes$ is the Kronecker product.
We set $q(x_t|x_{t-1})=f_t(x_t|x_{t-1})$
and we compare 
the RMSEs 
averaged over $1000$ MC runs.


\subsection{Variance of SR procedures}
\label{subsec:simu1}
\begin{figure}[htbp!]
\centering
\includegraphics[scale=0.4]{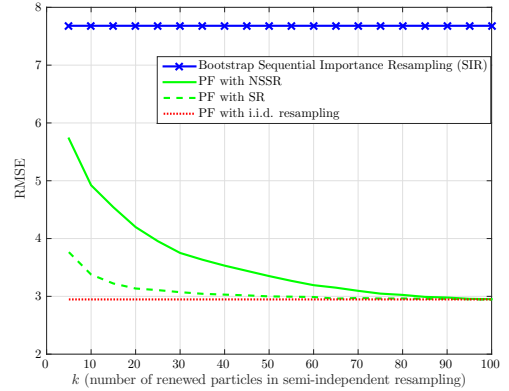}
\caption{\small RMSE as a function of $k$, tracking model.}
\label{fig:PolarTargetTrackingEMPVAR}
\end{figure}

We first analyze the behaviour of our algorithms
as a function of $k$. 
We set $N=100$, $\sigma_{\rho} = 0.1$ and $\sigma_{\theta} = \frac{\pi}{1800}$;
all MC runs use the same measurements.
Fig. \ref{fig:PolarTargetTrackingEMPVAR} 
displays the RMSE of $\widehat{\Theta}_t^{{\rm SR},N,k}$,
$\widehat{\Theta}_t^{{\rm NSSR},N,k}$,
$\widehat{\Theta}_t^{{\rm I-SIR},N}$ deduced from our resampling
schemes and $\widehat{\Theta}_t^{{\rm SIS},N}$ 
(a resampling step is computed at each time step but the estimate is taken before this step).
Of course, the performances of estimates based on the 
SR procedure 
improve when $k$ incrases. Even for small values of $k$, 
the improvement is significant.
It is also interesting to note that 
$\widehat{\Theta}_t^{{\rm SR},N,k}$ (resp.
$\widehat{\Theta}_t^{{\rm NSSR},N}$)
has the same performance as 
$\widehat{\Theta}_t^{{\rm I-SIR},N}$ when 
$k \geq N/2$ (resp. 
$k \geq 4N/5$).

\subsection{RMSE in the informative case at equal cost}
\label{subsec:simu2}
\begin{figure}[htb]
\centering
\includegraphics[scale=0.44]{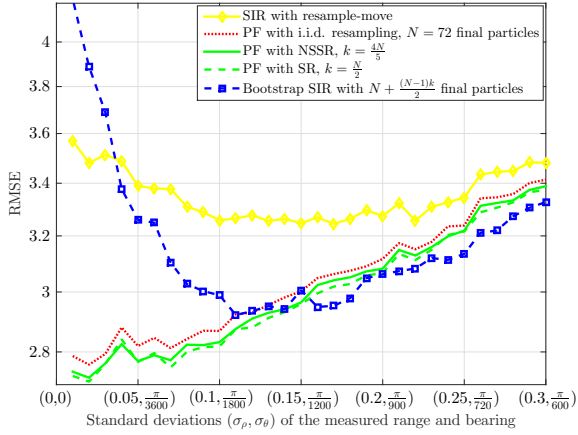}
\caption
{\small 
Tracking model, $\sigma_{\rho} \in [0.01,0.3]$ and $\sigma_{\theta} \in [\frac{\pi}{18000},\frac{\pi}{600}]$.
}
\label{fig:PolarTargetTrackingRMSEVaryMeasVar}
\end{figure}

We now compare our estimates 
with existing improvements of the PF in informative models. 
In particular, 
the PF with MCMC resample move is a popular solution
to introduce sample variety after resampling  \cite{berzuini-mcmc}.
Roughly speaking, the $N$ particles which follow the ({\textsl R.}) step 
of Algorithm \ref{algo-PFSIR} are moved via an MCMC algorithm 
with $k$ iterations (here an independent Metropolis-Hasting algorithm).
Thus, our SR procedure has the same computational 
cost in terms of sampling steps as the SIR PF with MCMC moves.
We also compare our estimates with those based on the classical SIR and I-SIR algorithms 
but with a given budget of total sampling (sampling + resampling) operations. 
We thus set $N=100$ particles and $k=N/2$
for the computation of $\widehat{\Theta}_t^{{\rm SR},N,k}$
and the estimate based on the resample move PF,
$N=72$ for that of $\widehat{\Theta}_t^{{\rm I-SIR},N}$
and $N+\frac{(N-1)k}{2}=2575$ particles for that of $\widehat{\Theta}_t^{{\rm SIS},N}$ . 
The global sampling cost for all these algorithms is approximately $(2N + Nk)$.
We also compute $\widehat{\Theta}_t^{{\rm NSSR},N,k}$
with $N=100$ and $k=4N/5$; 
its computation
does not have the same computational cost but can be parallelized.
The results are displayed in Fig. \ref{fig:PolarTargetTrackingRMSEVaryMeasVar}.

When the observations are very informative ($\sigma_\rho$ and $\sigma_\theta$ are small),
the classical solution tends to degenerate (it starts working
when $(\sigma_\rho,\sigma_\theta)= (0.15,\frac{\pi}{1200}))$, 
while our solutions are robust and present better performances.
As the variance of the measurement noise increases, 
the different estimates tend to behave similarly;
the classical SIR algorithm performs slightly better, 
which is not surprising since in this case it no longer suffers from the degeneracy
phenomenon 
and the number of final samples used is far superior to the other solutions.
We also observe that the resample move which uses differently the $k$ extra samples 
does not perform well when compared to the SR procedure in very informative models,
and is outperformed by our solutions when the observations are not informative.
Finally, 
our 
SR algorithm with $k = \frac{N}{2}$ outperforms the (totally)
independent resampling one when the budget is fixed.

\vspace{-0.5cm}

\section{Conclusion}

In this paper we revisited the resampling step of PF algorithms,
and proposed a resampling scheme where each new final particle 
is resampled from a support which is partially rejuvenated with $k$ new particles.
This yields a class of parameterized solutions 
which encompasses the classical multinomial resampling technique ($k=0$) 
and the independent resampling one ($k=N$),
enabling to tune the balance 
between variance and computational cost.
Simulations showed that choosing $k=N/2$ leads to similar performances
to the fully independent resampling procedure. 
Moreover, in very informative models our algorithm is not affected 
by the degeneration phenomenon,
contrary to the classical SIR algorithm.

\appendix [Proof of Proposition \ref{prop:1}]
\label{annex:proof1}

Let us consider a PF with resampling at time $t$. 
First, (\ref{prop:1:a}) holds because the SIR, I-SIR and SR procedures all produce resampled particles 
which, given $\{x_{0:t-1}^i\}_{i=1}^N$,
are (marginally) sampled from the same distribution $\tilde{q}^{N}$; 
and (\ref{prop:1:b}) is straightforward from (\ref{prop:1:c}) and the fact that 
SR reduces to SIR (resp. I-SIR) 
when $k = 0$ (resp. $k = N$). 
Let us address (\ref{prop:1:c}).
Since $\widehat{\Theta}_t^{{\rm SR},N,k}  = \frac{1}{N} \sum_{i=1}^N \varphi(x_t^{{\rm SR}, i})$,
given $\{x_{0:t-1}^i\}_{i=1}^N$ 
\begin{displaymath}
N^2 {\rm var}_k(\widehat\Theta_t^{{\rm SR},N,k}) 
\! = \!
\sum_{i=1}^{N} {\rm var}(\varphi(x_t^{i}))
+
2 \!\!\!\!\sum_{\substack{i_1,i_2 = 1 \\ i_1 < i_2}}^{N}
\!\!\!\!{\rm cov}_k(\varphi(x_t^{i_1}),\!\varphi(x_t^{i_2}));
\end{displaymath}
here index $k$ in a (co)variance emphasizes the fact that 
it depends on $k$.
The first term of the r.h.s. is independent of $k$ 
(and coincides with ${\rm var}(\widehat\Theta_t^{{\rm I-SIR},N})$),
so the difference between different values of $k$ stems from the covariance terms.
Next 
${\rm cov}_k(\varphi(x_t^{i_1}),\varphi(x_t^{i_2})) =$
${\rm E}_k[\varphi(x_t^{i_1})\varphi(x_t^{i_2})] -$ 
${\rm E}[\varphi(x_t^{i_1})]
 {\rm E}[\varphi(x_t^{i_2})]$,
and again, the second term of the r.h.s. is independent of $k$.
Finally for $i_1 < i_2$,

\begin{align*}
{\rm E}_k[\varphi(x_t^{i_1})\varphi(x_t^{i_2})] & = 
{\rm E}[{\rm E}[\varphi(x_t^{i_1})\varphi(x_t^{i_2})|\tilde{x_t}^{i_1:i_2,:}]] \\
& = {\rm E}[
{\rm E}[\varphi(x_t^{i_1})|\tilde{x}_t^{i_1,:}]
{\rm E}[\varphi(x_t^{i_2})|\tilde{x}_t^{i_2,:}]
] \\
& 
= {\rm E}[
\widehat{\Theta}^{\rm SIS}_{t}(\tilde{x}_t^{i_1,:})
\widehat{\Theta}^{\rm SIS}_{t}(\tilde{x}_t^{i_2,:})
] \\
& = {\rm E}[{\rm E}[
\widehat{\Theta}^{\rm SIS}_{t}(\tilde{x}_t^{i_1,:})
\widehat{\Theta}^{\rm SIS}_{t}(\tilde{x}_t^{i_2,:})
|m_{i_1+1}^{i_2}(1\mathalpha{:}k)]]
\end{align*}
where 
$m_{i_1+1}^{i_2}(1:k)$ 
represents all the indices redrawn 
from iterations $i_1 + 1$ to $i_2$
(the third equality holds because
$x_t^i$ is resampled from support $\tilde{x}_t^{i,:}$ 
(see Fig. \ref{fig:slices}),
so 
${\rm E}(\varphi(x_t^i)) = \widehat{\Theta}_t^{{\rm SIS}, N}(\tilde{x}_t^{i,:})$
where $\widehat{\Theta}_t^{{\rm SIS}, N}$ was defined in section \ref{classical-sir}).

The outer expectation in this last expression corresponds to a uniformly weighted sum 
over all possible values of $m_{i_1+1}^{i_2}(1\mathalpha{:}k)$, 
i.e. over $(A^{k}_{N})^{i_2 - i_1}$ terms 
where $A^{k}_{N}$ is the number of arrangements of $k$ among $N$. 
Given $m_{i_1+1}^{i_2}(1\mathalpha{:}k)$, 
the general term of this sum reads
\begin{flalign*}
& 
{\rm E}[
\widehat{\Theta}^{\rm SIS}_{t}(\tilde{x}_t^{i_1,:})
\widehat{\Theta}^{\rm SIS}_{t}(\tilde{x}_t^{i_2,:})
|m_{i_1+1}^{i_2}(1\mathalpha{\mathalpha{:}}k)]
= 
\\
& 
{\rm E}[
{\rm E}[
\widehat{\Theta}^{\rm SIS}_{t}(\tilde{x}_t^{i_1,:})
\widehat{\Theta}^{\rm SIS}_{t}(\tilde{x}_t^{i_2,:})
|\tilde{x}_t^{i_1,1:N \textbackslash m_{i_1+1}^{i_2}(1\mathalpha{:}k)} 
]
|m_{i_1+1}^{i_2}(1\mathalpha{:}k)] 
\end{flalign*}
where $\tilde{x}^{i_1,1:N \textbackslash m_{i_1+1}^{i_2}(1\mathalpha{:}k)}$ 
are the particles shared by supports $\tilde{x}^{i_1,:}$ and $\tilde{x}^{i_2,:}$. 
Under this conditioning, 
$\widehat{\Theta}^{\rm SIS}_{t}(\tilde{x}_t^{i_1,:})$ and 
$\widehat{\Theta}^{\rm SIS}_{t}(\tilde{x}_t^{i_2,:})$ 
are independent so the general term is
\begin{align*}
{\rm E}[ & 
{\rm E}[
\widehat{\Theta}^{\rm SIS}_{t}(\tilde{x}_t^{i_1,:})
|\tilde{x}_t^{i_1,1:N \textbackslash m_{i_1+1}^{i_2}(1\mathalpha{:}k)} 
]
 \\
& 
\times 
{\rm E}[
\widehat{\Theta}^{\rm SIS}_{t}(\tilde{x}_t^{i_2,:})
|\tilde{x}_t^{i_1,1:N \textbackslash m_{i_1+1}^{i_2}(1\mathalpha{:}k)} 
]
|
m_{i_1+1}^{i_2}(1\mathalpha{:}k)] 
\\
& = 
{\rm E}[ 
{\rm E}^2[
\widehat{\Theta}^{\rm SIS}_{t}(\tilde{x}_t^{i_1,:})
|\tilde{x}^{i_1,1:N \textbackslash m_{i_1+1}^{i_2}(1\mathalpha{:}k)} 
]
|
m_{i_1+1}^{i_2}(1\mathalpha{:}k)] \\
&= h(m_{i_1+1}^{i_2}(1\mathalpha{:}k))=h(m_{i_1+1}^{i_2}(1\mathalpha{:}k-1),m_{i_1+1}^{i_2}(k))
\end{align*}
because given the trajectories from the previous time steps,
particles from different supports are all marginally drawn from the same densities.
Finally

\begin{align}
\label{expression}
& \! \! {\rm E}_k(\varphi(x_t^{i_1})  \varphi(x_t^{i_2})) = 
\frac{1}{(A^{k}_{N})^{i_2-i_1}}  \!\!\!
\sum\limits_{m_{i_1+1}^{i_2}(1\mathalpha{:}k)} \!\!\!\! 
h(m_{i_1+1}^{i_2}(1\mathalpha{:}k)) \text{.}
\end{align}

It remains to compare \eqref{expression} with
the same expression with $k \leftarrow k-1$.
We observe that \eqref{expression} can be rewritten 
as 
\begin{align*}
{\rm E}_k&(\varphi(x_t^{i_1})  \varphi(x_t^{i_2})) =
\frac{1}{(A^{k-1}_{N})^{i_2-i_1}} 
\sum\limits_{m_{i_1+1}^{i_2}(1\mathalpha{:}k-1)} \\ &
\frac{1}{N-k+1} \sum\limits_{m_{i_1+1}^{i_2}(k)}
h(m_{i_1+1}^{i_2}(1\mathalpha{:}k-1),m_{i_1+1}^{i_2}(k)) \text{,}
\end{align*}
where the second line coincides with the conditionnal expectation
${\rm E}[h(m_{i_1+1}^{i_2}(1\mathalpha{:}k))|m_{i_1+1}^{i_2}(1\mathalpha{:}k\mathalpha{-}1)]$. 
Given $m_{i_1+1}^{i_2}(1 \mathalpha{:} k-1)$, the set \! \!
$\tilde{x}^{i_1,1:N \textbackslash m_{i_1+1}^{i_2}(1\mathalpha{:}k)}$
is included in $\tilde{x}^{i_1,1:N \textbackslash m_{i_1+1}^{i_2}(1\mathalpha{:}k-1)}$;
consequently, the Rao-Blackwell decomposition
(${\rm E}({\rm E}^2(X|Y)) \leq  {\rm E}({\rm E}^2(X|Y,Z))$)
ensures that
\begin{align*}
h(m_{i_1+1}^{i_2}(1\mathalpha{:}k-1),m_{i_1+1}^{i_2}(k)) \leq h(m_{i_1+1}^{i_2}(1\mathalpha{:}k-1))
\end{align*}
for all $m_{i_1+1}^{i_2}(k)$,
and so that ${\rm E}[h(m_{i_1+1}^{i_2}(1\mathalpha{:}k))|m_{i_1+1}^{i_2}(1\mathalpha{:}k-1)]$ 
$\leq$ $h(m_{i_1+1}^{i_2}(1\mathalpha{:}k-1))$, whence (\ref{prop:1:c}).

\bibliographystyle{ieeetr}
\bibliography{bib-yohan}

\end{document}